\documentclass[12pt]{article}
\usepackage{graphicx}
\usepackage{psfrag}
\usepackage{epstopdf}
\usepackage{latexsym}
\usepackage{slashed}

\begin{document}

\begin{titlepage}
\null\vspace{-62pt}

\pagestyle{empty}
\begin{center}

\vspace{1.0truein} {\Large\bf Gauge independence of the physical fermion mass in the effective action of theories with radiative symmetry breaking}

\vspace{1in}
{\large Dimitrios Metaxas} \\
\vskip .4in
{\it Department of Physics,\\
National Technical University of Athens,\\
Zografou Campus, GR 15773 Athens, Greece\\
 metaxas@central.ntua.gr}\\

\vspace{.5in}
\centerline{\bf Abstract}

\baselineskip 18pt
\end{center}
I consider an Abelian gauge theory, with a complex scalar field and massless, chiral fermions, in the limit where the couplings and the interactions induce radiative symmetry breaking,
and I derive and verify the Nielsen identities that ensure the gauge independence of the physical fermion mass that emerges in the broken phase of the effective action.

\end{titlepage}
\newpage
\pagestyle{plain}
\setcounter{page}{1}
\newpage

\section{Introduction}

The phenomenon of symmetry breaking is fundamental in the physics of the Standard Model, which is consistently formulated as an interacting theory of massless fermions
and gauge bosons, that acquire mass in the broken phase of the Higgs sector.

The gauge freedom present in the original action of the Standard Model, as well as any other gauge theory of particle physics, makes the calculations of additional radiative corrections more involved, by the process of gauge fixing and the 
resulting gauge dependence that is present in the calculation of individual Feynman diagrams; the final result, however, for any physical quantity, calculated at a given order
of the coupling constants, is expected to be gauge independent once all the relevant terms are accounted for.

The gauge dependence of individual terms is described by the Nielsen identities \cite{nielsen1}, that are expected to hold for the generating functionals of the Green functions of such theories,
and have been used in order to demonstrate the gauge independence of physical quantities such as masses, tunneling rates, and other consequences of the phenomenon of
symmetry breaking in particle physics and cosmology \cite{nielsen2, nielsen3}.

In this work, I will consider the specific problem of the gauge independence of the fermion masses that are generated for originally massless, chiral fermions, in the process
of symmetry breaking, in the limiting case where the latter is induced radiatively instead of being present in the tree-level action \cite{cw}.
Although some comprehensive diagrammatic treatments of similar problems, in the Standard Model and other cases, exist in the literature, 
using the Nielsen identities or other methods \cite{sm}, here, I will consider the calculation and definition of particle mass through the effective action, 
and derive and prove the relevant Nielsen identities at this level. 
This is important for the consistency of the theory, the power-counting and collection of the relevant terms of a physical calculation,
as well as for the use of the effective action formalism as a tool for the study of the process and consequences of symmetry breaking.

In Sec.~2, I describe the general formalism of the Nielsen identities for the effective action and  the model and parameters considered.

In Sec.~3, I
derive the Nielsen identities that ensure the gauge independence of the physical fermion mass in the effective action and give an 
 explicit verification in terms of an expansion in the coupling constant in the case of radiative symmetry breaking. In Sec.~4, I conclude with some comments.

\section{The general formalism}

For an Abelian gauge theory with fields denoted collectively by $\phi_i$ and transforming as 
\begin{equation}
\delta\phi_i = \Delta\phi_i \,\theta,
\end{equation}
under an infinitesimal gauge tranformation with a gauge parameter $\theta$,
the classical, tree-level action, $S_0$, has to be gauge-fixed by choosing a gauge condition, $F(\phi_i)$, and 
adding the terms
\begin{equation}
-\frac{F(\phi_i)^2}{2 \xi} - \bar{\eta} \frac{\delta F}{\delta \phi_i} \Delta\phi_i \, \eta,
\end{equation}
with Fadeev-Popov ghosts $\eta$ and $\bar{\eta}$,
rendering the various Green functions dependent on the gauge parameter $\xi$.

The Nielsen identities \cite{nielsen1, nielsen2} describe the dependence of the effective action, $S$  (the generating functional of the
one-particle irreducible Green functions), as
\begin{equation}
\xi \frac{\partial S}{\partial \xi} = \int \frac{\delta S}{\delta \phi_i} H_i[\phi_j],
\label{n1}
\end{equation}
where the functional
\begin{equation}
H_i[\phi_j] =\-\frac{i}{2}  < \Delta\phi_i \,\eta \bar{\eta}\, F >,
\label{n2}
\end{equation}
is a sum of the respective one-particle irreducible graphs with the described operator insertions, and can also be expanded in the fields $\phi_j$ and their derivatives.

Here, I will consider the theory of  an Abelian gauge field, interacting with a complex scalar field, as well as massless chiral fernions with Yukawa interactions,
and general hypercharges for the gauge transformations of the scalar and fermion fields.

The Lagrangian 
\begin{eqnarray}
{\cal L} = -\frac{1}{4} \, F_{\mu\nu}^2 + (D_\mu \Phi)^* (D_\mu \Phi) - U_0(\Phi) + \\ \nonumber
+ \bar{\psi}_L i \slashed{D} \psi_L +\bar{\psi}_R i \slashed{D} \psi_R  + \\ \nonumber
-\sqrt{2} g (\Phi\, \bar{\psi}_L \psi_R + \Phi^* \bar{\psi}_R \psi_L)
\end{eqnarray} 
is written, with the metric tensor $\eta^{\mu\nu} = {\rm diag}(+---)$,  in terms of the gauge field $A_\mu$, with the gauge field strength $F_{\mu\nu}=\partial_\mu A_\nu - \partial_\nu A_\mu$,
the complex scalar field $ \Phi = \frac{1}{\sqrt{2}} (\Phi_1 + i \Phi_2)$ and the left- and right-handed chiral fermion fields, $\psi_L$ and $\psi_R$.

The matter fields are charged under the gauge interaction, with the respective hypercharges, $Y$, $Y_L$ and $Y_R$  appearing in the covariant derivatives
\begin{equation}
D_\mu \Phi = (\partial_\mu + ie Y A_\mu) \Phi,
\end{equation}
\begin{equation}
D_\mu \psi_L = ({\partial_\mu} + ie Y_L {A_\mu}) \psi_L,
\end{equation}
\begin{equation}
D_\mu \psi_R = ({\partial_\mu} + ie Y_R {A_\mu}) \psi_L,
\end{equation}
with the coupling constant, $e$,
and the hypercharges that satisfy
\begin{equation}
Y=Y_L - Y_R,
\label{y1}
\end{equation}
by the requirement of gauge invariance.

I note in passing, that the theory, as written, has a chiral anomaly which, however, can be canceled, similarly to 
the case of the Standard Model, in the presence of more chiral fermion ``generations", when
the relation 
$\sum_L Y_L^3 -\sum_R Y_R^3 =0$  is satisfied. This can be trivially done here, and will not affect the results, so it will not be explicitly 
shown.

The fermion algebra can be described by the Dirac matrices, $\gamma^\mu$, that satisfy the anticommutation relations
\begin{equation}
\{ \gamma^\mu, \gamma^\nu \} = 2 \, \eta^{\mu\nu},
\end{equation}
as well as $\gamma^5$, with $\{ \gamma^\mu, \gamma^5 \} =0$, $(\gamma^5)^2=1$, $(\gamma^5)^\dagger=\gamma^5$, 
and the projectors
$P_{ R,L} =\frac{1\pm\gamma^5}{2}$, with $P_{ R,L}^2 = P_{ R,L}$, $P_{ L} +P_{ R} =1$, $P_L \, P_R =0$.
Also,
the usual notation for spinors is
$\bar{\psi} = \psi^\dagger \gamma^0$,
and the Dirac slash  is $\slashed{a}=\gamma^\mu \, a_\mu$, for any vector quantity $a_\mu$. Where implied, the unit matrix is simply denoted by $1$.

Then, the chiral spinors can be combined in a Dirac spinor,
$\psi=\psi_L +\psi_R$, 
with $\psi_L=P_L \, \psi$ and $\psi_R =P_R \, \psi$,
and the Lagrangian can be written as
\begin{eqnarray}
{\cal L} = -\frac{1}{4} \, F_{\mu\nu}^2 + (D_\mu \Phi)^* (D_\mu \Phi) - U_0(\Phi) + \\ \nonumber
+ \bar{\psi}\,  i [\slashed{\partial} + ie \slashed{A} (V-A \gamma^5)]  \psi
- g  \bar{\psi} (\Phi_1 + i \gamma^5 \Phi_2) \psi,
\end{eqnarray} 
with
\begin{equation}
V=\frac{Y_L+Y_R}{2}\,,\,\,\,A=\frac{Y_L-Y_R}{2},
\label{y2}
\end{equation}
in order to derive the Feynman rules for the fermion propagators and interactions.

The gauge invariance of the theory is expressed in the infinitesimal gauge transformations
\begin{eqnarray}
\delta A_\mu &=&\partial_\mu \theta  \\
\delta \Phi &=& - i e Y \theta\, \Phi  \\
\delta \psi_L &=&- i e Y_L\, \theta\, \psi_L  \\
\delta \psi_R &=& - i e Y_R\, \theta\, \psi_R ,
\end{eqnarray}
that also  imply
\begin{eqnarray}
\delta \Phi_1 &=& e Y \theta\, \Phi_2 \\
\delta \Phi_2 &=& - e Y \theta\, \Phi_1 \\
\delta \psi &=& - i e (V- A \gamma^5) \,\theta\,\psi,
\end{eqnarray}
and can be used to extract the $\Delta_i$'s for the various fields.

The tree-level potential for the scalar field is given  by
\begin{equation}
U_0(\Phi) = m_0^2\, \Phi^* \Phi +\frac{\lambda}{6} (\Phi^* \Phi)^2,
\end{equation}
and, if the mass parameter $m_0^2$ is positive, it has a symmetric minimum at $\Phi=0$.
However, one-loop effects in the effective action, give a non-zero, symmetry breaking vacuum at $<\Phi> = \mu$,
when the coupling parameter $\lambda$ is of order $e^4$, and $m_0^2$ is also sufficiently small, of order $e^2 \mu^2$ (or zero) \cite{cw}.

In order to study this interesting phenomenon, it is useful to study the effective action, which can be obtained by the one-particle irreducible
graphs after making the shift $\Phi_1 \rightarrow \Phi_1 + \phi$ and dropping the terms linear in the fields, in order to derive the
$\phi$-dependent Feynman rules. Then the effective action can be derived in a coupling constant and derivative expansion in  $\phi$,
and the emergence of a symmetry-breaking minimum can be explored.

Finally, in order to derive the full set of  Feynman rules for the  theory,
with the gauge-fixing and ghost terms, as described in the beginning of this Section,
 the class of generalised $R_\xi$ gauges will be used, with the condition
\begin{equation}
F = \partial_\mu A^\mu + e Y v\, \Phi_2,
\end{equation}
which includes an additional gauge parameter, $v$.
Usually, the choice $v=-\xi <\Phi>$ is made, in order to eliminate mixing terms in the propagators,
since, however, the $\xi$-dependence of $<\Phi>$ will also be involved, it will be convenient to use
this general gauge-fixing condition, with an arbitrary value for $v$,  although other choices are also possible \cite{nielsen2, nielsen3, sm}.
An obvious drawback is that an additional gauge parameter is used, and the final result has to be 
$v$-independent as well as $\xi$-independent; the relevant procedure, however, is identical to the one done for  the $\xi$-dependence,
and can be easily performed \cite{nielsen2}.

\section{Verification of the Nielsen identities}

The general formalism described in the previous Section gives, for the specific model used, an effective action
with the general form
\begin{eqnarray}
S&=&\int Z (\phi)\, \frac{1}{2}\, (\partial \phi)^2 - U(\phi) + \\ \nonumber
&+& Z_L(\phi)\, \bar{\psi}_L \, i \slashed{\partial} \,\psi_L + Z_R(\phi)\, \bar{\psi}_R\, i \slashed{\partial}\, \psi_R - \\ \nonumber
&-& \tilde{Z} (\phi)\, (\bar{\psi}_L \,{\psi}_R + \bar{\psi}_R \,\psi_L),
\end{eqnarray}
where all the $Z$'s are $\xi$-dependent beyond their tree-level value,
which, for $Z$, $Z_L$ and $Z_R$ is equal to one, and for $\tilde{Z}$ it is equal to $g\, \phi$
(usually, I will drop the explicit $\phi$-dependence where obvious).

In order to apply the Nielsen identity (\ref{n1}), the functionals in (\ref{n2}) need to be also expanded
in powers of derivatives and coupling constants in the scalar and fermion fields.
The lower order terms in the expansion can be written out in
\begin{equation}
\xi \frac{\partial S}{\partial \xi} =\int \frac{\delta S}{\delta \phi} \, C(\phi) + \frac{\delta S}{\delta \psi_L} C_L(\phi) \, \psi_L
+\frac{\delta S}{\delta\psi_R} C_R(\phi) \, \psi_R + c.c.,
\end{equation}
and the respective powers and derivative terms can be matched in the resulting expression, a procedure
which, after some straightforward algebra, gives the identities
\begin{equation}
\xi \frac{\partial{U}}{\partial{\xi}} = C\, \frac{\partial U}{\partial \phi}
\label{id1}
\end{equation}
\begin{equation}
\frac{1}{2} \xi \frac{\partial Z}{\partial \xi} = \frac{1}{2} C\frac{\partial Z}{\partial \phi} + Z \frac {\partial C}{\partial \phi}
\label{id2}
\end{equation}
\begin{equation}
\xi \frac{\partial Z_L}{\partial \xi} = C \frac{\partial Z_L}{\partial \phi}  +2\, C_L  \, Z_L
\label{id3}
\end{equation}
\begin{equation}
\xi \frac{\partial Z_R}{\partial \xi} = C \frac{\partial Z_R}{\partial \phi}  +2\, C_R  \, Z_R
\label{id4}
\end{equation}
\begin{equation}
\xi \frac{\partial \tilde{Z}}{\partial \xi} = C \frac{\partial\tilde{Z}}{\partial \phi}  + (C_L + C_R) \tilde{Z}.
\label{id5}
\end{equation}

The first identity (\ref{id1}) is the original Nielsen identity \cite{nielsen1}, that was derived and used in order to show
the gauge independence of the phenomenon of symmetry breaking through the effective action, by compensating the 
$\xi$-dependence of the effective potential with a respective field dependence 
\begin{equation}
\xi \frac{\partial\phi}{\partial \xi} = - C(\phi).
\label{id6}
\end{equation}

The second identity (\ref{id2}) was derived in \cite{nielsen2} in order to demonstrate the gauge independence of the vacuum decay rate
in theories with radiative symmetry breaking. It can also be used to show the gauge independence of the physical scalar mass
in the symmetry breaking minimum of the effective potential (a fact that was already derived before diagrammatically)
defined by
\begin{equation}
m^2_\phi=\frac{U''}{Z}  \,\,\,\, {\rm at }\,\,\, \,U'=0
\end{equation}
(primes will generally denote derivatives with respect to $\phi$).
Specifically, the relation
\begin{equation}
\xi\frac{\partial}{\partial\xi} m^2_\phi = C\frac{\partial}{\partial\phi} m^2_\phi
\end{equation}
can also be easily derived from (\ref{id1}) and (\ref{id2}) and their derivatives, evaluated at $U'=0$.

In this work, the gauge independence of the physical fermion mass,
defined as
\begin{equation}
m_\psi = \frac{\tilde{Z}}{\sqrt{Z_L\,Z_R}},
\end{equation}
will be demonstrated, after verifying the identities (\ref{id3}), (\ref{id4}) and (\ref{id5}),
which also easily lead to 
\begin{equation}
\xi\frac{\partial}{\partial\xi} m_\psi = C\frac{\partial}{\partial\phi} m_\psi.
\label{idfermion}
\end{equation}
Interestingly, (\ref{idfermion}) holds generally and not just at the minimum of the effective potential, although its
interpretation as a physical particle mass is naturally made at the broken phase.

In order to explicitly verify the new identities, the Feynman diagrams contributing to the various factors will be calculated
for the theory described in the previous Section, in the first non-trivial order of expansion in the coupling constant, $e$
(our assumption of radiative symmetry breaking implies that the other coupling constant, $\lambda$, is of order $e^4$).
Since we are interested specifically in the $\xi$-dependence, the propagators that will contribute
include the $\Phi_2$  and longitudinal $A_\mu$ propagators,
\begin{equation}
G_2 (k) = \frac{i (k^2 - \xi e^2 Y^2 \phi^2)}{D(k)},
\end{equation}
\begin{equation}
G_{\mu\nu}^L(k)=-i\frac{\xi(k^2-m_2^2) - e^2 Y^2 v^2}{D(k)} \frac{k_\mu k_\nu}{k^2},
\end{equation}
the ghost propagator,
\begin{equation}
\frac{i}{k^2+e^2 Y^2 v \phi},
\end{equation}
and the mixed $\Phi_2 - A_\mu$ propagator,
\begin{equation}
G_{2\mu} = \frac{ eY(\xi\phi+v)k_\mu}{D(k)}
\end{equation}
(where the  momentum is directed from $\Phi_2$ to $A_\mu$).

The fermion propagator is also ``dressed'' with the scalar field,
\begin{equation}
G_{\psi \bar{\psi}}=i \frac{\slashed{k}+g \phi}{k^2 - g^2\phi^2},
\end{equation}
and the denominators include the function
\begin{equation}
D(k) = k^4 - k^2 (m_2^2 - 2 e^2Y^2 v \phi) +e^2 Y^2 \phi^2 (e^2 Y^2 v^2 +\xi m_2^2),
\end{equation}
where $m_2^2=m_2^2(\phi)$ is the mass of the Goldstone boson. Normally, the relation
$\phi \, m_2^2(\phi) = U_0'(\phi)$ would hold, however, since we consider the limit of radiative symmetry breaking, which includes
the resummation of the transverse photon loops that are of the same order of magnitude as the tree level potential, $\lambda\sim e^4$,
and they modify it accordingly, we also have the relation $\phi \, m_2^2(\phi) = U'(\phi)$, where $U(\phi)$ is the effective potential,
for the Goldstone boson mass. 
This is used in the verification of the first Nielsen identity (\ref{id1}), however,
since $U$ and $m_2^2$ are of order $e^4$, the $\xi$-dependent terms arising from the denominators will be subleading
for most terms contributing to the $Z$-factors, for which the expansion starts at $e^2$ for the first non-trivial terms that will be calculated here.

The $C$ factors in the Nielsen identities are obtained from the diagrams in Fig.~1 and Fig.~2. The former contains the diagrams for the
calculation of $C$, the first term in the expansion of $-\frac{i}{2}<\Delta \Phi_1 \eta \bar{\eta} F >$, and the latter
the diagrams for the calculation of 
\begin{equation}
C_\psi = -\frac{i}{2} < \Delta\psi \eta \bar{\eta} F> = C_L \psi_L +  C_R \psi_R = C_1 \psi + C_2 \gamma^5 \psi,
\end{equation}
with $C_L=C_1 - C_2$, $C_R=C_1 +C_2$,
from which the relevant factors can be extracted.

The $Z$'s can be calculated from the diagrams of Fig.~3. Since we use ``dressed'' propagators, the
wave-function renormalization factors, $Z_L$ and $Z_R$ can be extracted from the terms dependent on the external fermion momentum, 
after writing
\begin{equation}
Z_L \bar{\psi_L}\, i \slashed{\partial} \psi_L + Z_R \bar{\psi_R}\,  i \slashed{\partial} \psi_R =
Z_1 \,\bar{\psi}\, i \slashed{\partial} \psi + Z_2 \, \bar{\psi}\, i \slashed{\partial} \gamma^5 \psi,
\end{equation}
with $Z_L=Z_1-Z_2$, $Z_R=Z_1+Z_2$,  and  the vertex factor, $\tilde{Z}$, can also be calculated from the momentum independent terms
of the same diagrams.

In the Figures, the wavy lines denote the gauge field, $A_\mu$, the dashed lines denote the $\Phi_2$ field, 
double solid lines denote the fermion field, $\psi$, and the dotted lines the ghosts.
The blobs in Fig.~1 denote the insertions of $F$ and $\Delta\Phi_1$ and in Fig.~2 the insertions of $F$ and $\Delta\psi$.

The $C$ and $C_{L,R}$ factors turn out to be of order $e^2$ in the leading terms, while $Z_{L,R}$ have an expansion of the form
$1+e^2...$, with the $\xi$-dependence in the $e^2$ terms. The corresponding expansion for $\tilde{Z}$ is of the form
$\tilde{Z}=g\,\phi\,(1+\tilde{Z}_\xi)$, with $\tilde{Z}_\xi$ also of order $e^2$ in the leading, $\xi$-dependent, term.

Thus, the required Nielsen identities, that describe the gauge independence of the physical fermion mass (\ref{id3}, \ref{id4}, \ref{id5})
become
\begin{equation}
\xi\frac{\partial Z_L}{\partial \xi} = 2 C_L,
\label{nid3}
\end{equation}
\begin{equation}
\xi\frac{\partial Z_R}{\partial \xi} = 2 C_R,
\label{nid4}
\end{equation}
\begin{equation}
\xi\frac{\partial \tilde{Z}_\xi}{\partial \xi} =  \frac{C}{\phi}+ (C_L+C_R),
\label{nid5}
\end{equation}
and can be verified in an explicit calculation of the Feynman diagrams in the Figures. The $C$ factor \cite{nielsen1, nielsen2} is calculated
from the sum of the diagrams in Fig.~1 as
\begin{eqnarray}
\nonumber C(\phi) &=& -\frac{i}{2} eY\int_k \frac{eY(\xi\phi +v) k^2 - eYv(k^2-\xi e^2Y^2\phi^2)}{(k^2+e^2 Y^2 v \phi) D(k)} = \\
            &=&-\frac{i}{2}\,\xi\, e^2 \,Y^2\, \phi \, I,
\label{cres}
\end{eqnarray}
where $\int_k$ denotes the integration $\int d^4k /(2\pi)^4$ over the loop momentum, $k$, and
\begin{equation}
I=I(\phi, \xi)=\int_k \frac{1}{D(k)}
\end{equation}
is a suitably regularised and renormalised logarithmic function of the parameters of the theory, as described before.

The $C_{L,R}$ factors can be obtained from the sum of the diagrams in Fig.~2 that gives
\begin{eqnarray}
\nonumber C_\psi&= &-\frac{ie^2}{2}\int_k \frac{(\xi k^2 -e^2Y^2v^2)+e^2Y^2(\xi\phi v+v^2)}{D(k) (k^2 +e^2 Y^2 v\phi)}
(V-A \gamma^5)^2 = \\
&=&-\frac{i e^2 \xi}{2} I\,(V^2+A^2 -2 V A \gamma^5),
\end{eqnarray}
and consequently the factors $C_{1,2}$, as explained before.

The $Z$-factors are obtained from the diagrams of Fig.~3, with external momentum $p$ in the fermion lines.
The first one gives
\begin{equation}
\xi e^2 \, I (V+A \gamma^5)(\slashed{p}- g\,\phi)(V-A \gamma^5),
\end{equation}
while the sum of the other two becomes 
\begin{equation}
-g\,\phi \,\xi \,e^2 \, 2 A Y I,
\end{equation}
which gives the results
\begin{equation}
Z_1 =- i e^2 \xi I (V^2 +A^2)\,\,,\,\,Z_2 = i e^2 \xi \, I \,2 VA,
\end{equation}
and 
\begin{equation}
\tilde{Z}_\xi=-i \xi e^2 \, I\, (V^2 - A^2 +2 AY)
\end{equation}

Collecting the various factors from the previous relations, and using (\ref{y1}) and (\ref{y2}), the relations
(\ref{nid3}), (\ref{nid4}) and (\ref{nid5}) for the Nielsen identities are verified.

\section{Comments}

In this work, the extension of the Nielsen identities for the effective action has been described, in theories that involve chiral fermions,
that acquire mass through the mechanism of radiative symmetry breaking. This has been done with a general ``hypercharge'' assignment, and is expected to
be similar for theories like the Standard Model, validating the use of the effective action formalism in order to extract gauge independent physical quantities,
like the fermion masses considered here. Similar work can be done in models involved in cosmology and particle physics phenomenology, in zero and finite temperature,
as well as curved spacetime backgrounds.

\newpage

\begin{figure}
\centering
\includegraphics[width=60mm]{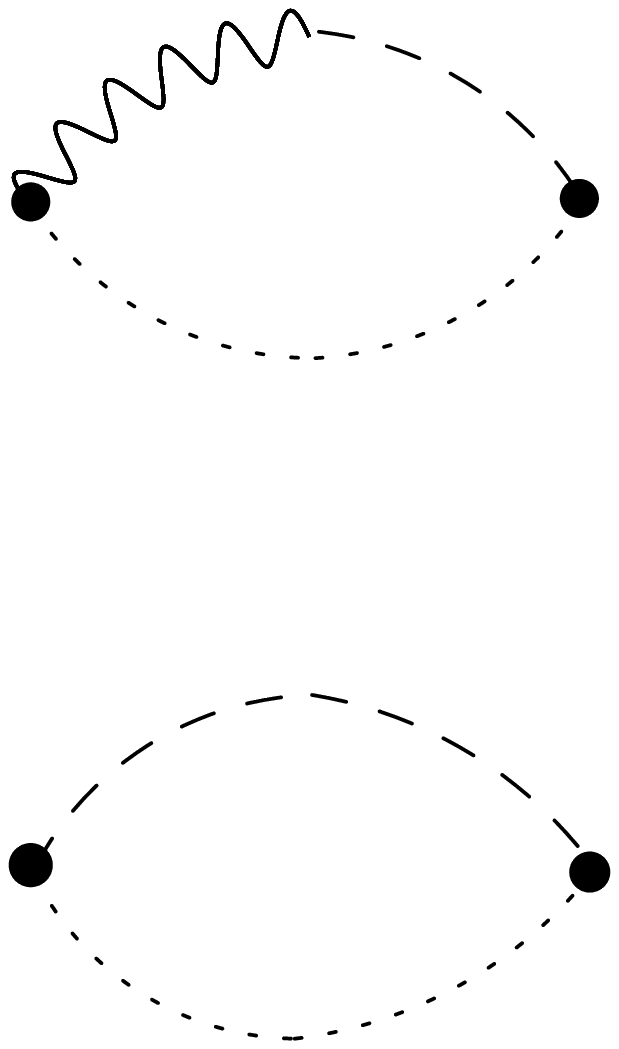}

\caption{   The diagrams involved in the calculation of $C(\phi)$.    }
\end{figure}

\begin{figure}
\centering
\includegraphics[width=60mm]{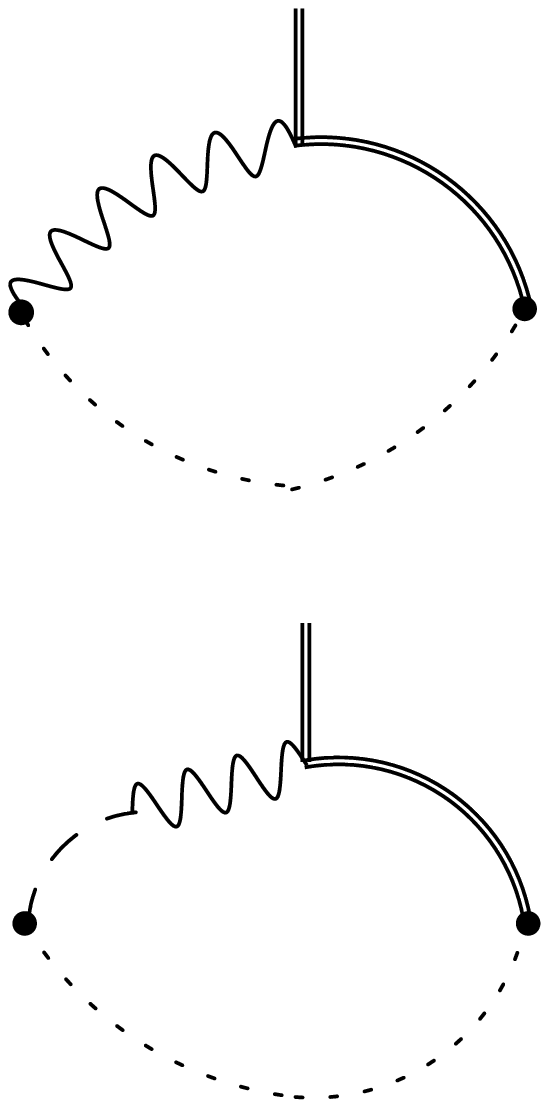}

\caption{   The diagrams involved in the calculation of $C_\psi$.    }
\end{figure}

\begin{figure}
\centering
\includegraphics[width=60mm]{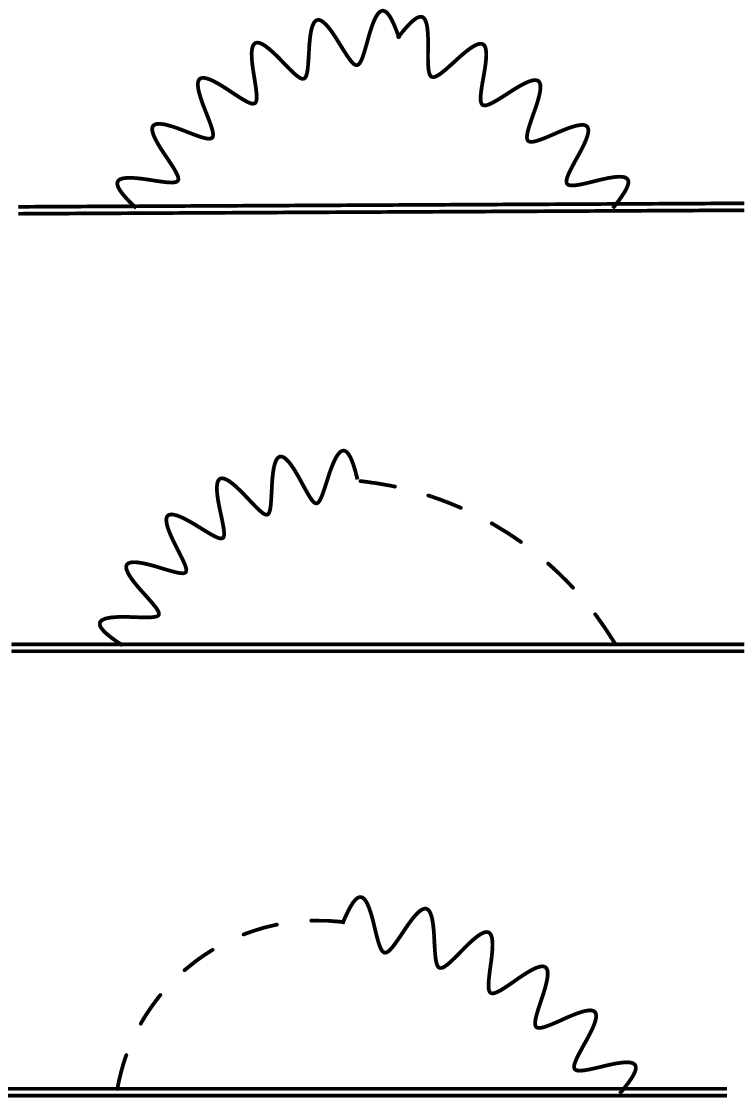}

\caption{  The diagrams involved in the calculation of $Z_{L,R}$ and $\tilde{Z}$.     }
\end{figure}


\vspace{0.5in}




\begin{thebibliography}{99}

\bibitem{nielsen1}  N.~K.~Nielsen, {\it Nucl. Phys.} {\bf B101}, 173 (1975).

                             R.~Fukuda and T.~Kugo, {\it Phys. Rev.} {\bf D13}, 3469 (1976).

                           R.~Kobes, G.~Kunstatter and A.~Rebhan, {\it Nucl. Phys.} {\bf B355}, 1 (1991).

                          I.~J.~R.~Aitchison and C.~M.~Fraser, {\it Ann. Phys. (N.Y.)} {\bf 156}, 1 (1984).              

\bibitem{nielsen2} D.~Metaxas and E.~J.~Weinberg, {\it Phys. Rev.} {\bf D53}, 836 (1996).

                          D.~Boyanovsky, W.~Loinaz and R.~S.~Willey, {\it Phys. Rev.} {\bf D57}, 100 (1998).

                         J.~Baacke and K.~Heitmann, {\it Phys. Rev.} {\bf D60}, 105037 (1999).

                       D.~Metaxas, {\it Phys. Rev.} {\bf D63} 085009 (2001).

                    L.~P.~Alexander and A.~Pilaftsis, {\it J.Phys.} {\bf G36}, 045006 (2009).

                C.~L.~Wainwright, S.~Profumo and J.~Ramsey-Musolf, {\it Phys. Rev.} {\bf D86}, 083537 (2012).

               M.~Garny and T.~Konstandin, {\it JHEP} {\bf 07}, 189 (2012).

\bibitem{nielsen3}  S.~P.~Martin, {\it Phys. Rev.} {\bf D90}, 016013 (2014).

                          A.~Andreassen, W.~Frost and M.~D.~Schwartz, {\it Phys. Rev.} {\bf D91}, 016009 (2015).

                       L.~Di~Luzio, G.~Isidori and G.~Ridolfi, {\it Phys. Lett} {\bf B753}, 150 (2016).

                     A.~D.~Plascencia and C.~Tamarit, {\it JHEP} {\bf 10}, 099 (2016).

                    Z.~Lalak, M.~Lewicki and P.~Olszewski, {\it Phys. Rev.} {\bf D94}, 085028 (2016).

                  J.~R.~Espinosa, M.~Garny and T.~Konstandin, {\it Phys. Rev.} {\bf D94}, 055026 (2016).
                         
                     J.~R.~Espinosa, M.~Garny, T.~Konstandin and A.~Riotto, {\it Phys. Rev.} {\bf D95}, 056004 (2017).

                  L.~Chataignier, T.~Prokopec, M.~G.~Schmidt and B.~Swiezewska, {\it JHEP}  {\bf 08}, 083 (2018).

                         A.~Urbano, {\it JCAP} {\bf 04}, 040 (2020).

\bibitem{cw}   S.~R.~Coleman and E.~J.~Weinberg, {\it Phys. Rev.} {\bf D7}, 1888 (1973).

                       E.~J.~Weinberg, {\it Phys. Rev.} {\bf D47}, 4614 (1993).

\bibitem{sm}  P.~Gambino and P.~A.~Grassi, {\it Phys. Rev.} {\bf D62}, 076002 (2000).

A.~Freitas and D.~Stockinger, {\it Phys. Rev.} {\bf D66}, 095014 (2002).

S.~P.~Martin, {\it Phys. Rev.} {\bf D71}, 116004 (2005).

D.~Binosi and J.~Papavassiliou, {\it JHEP} {\bf 03}, 041 (2007).

M.~Krause, R.~Lorenz, M.~Muhlleitner, R.~Santos and H.~Ziesche, {\it JHEP} {\bf 09}, 143 (2016).

N.~Irges and F.~Koutroulis, {\it Nucl. Phys.} {\bf B924}, 178 (2017).

A.~Denner, L.~Jenniches, J.~Lang and C.~Sturm, {\it JHEP} {\bf 09}, 115 (2016).

M.~A.~L.~Capri, D.~Dudal, A.~D.~Pereira, D.~Fiorentini and M.~S.~Guimaraes, {\it Phys. Rev.} {\bf D95}, 045011 (2017).


\end{thebibliography}
\end{document}